\documentclass[a4paper]{jpconf}
\usepackage{graphicx}
\usepackage{enumitem}
\usepackage{enumerate}
\usepackage[numbers]{natbib}
\usepackage{enumitem}
\usepackage{enumerate}
\usepackage[font={footnotesize}]{caption}
\usepackage{url}
\usepackage{scalerel}
\usepackage{setspace}
\usepackage{enumitem}
\usepackage[version=3]{mhchem}

\singlespace
\usepackage{caption}

\def\apj{ApJ}
\def\apjl{ApJ Let.}

\def\aap{A\&A}
\def\aap{A\&A}
\def\mnras{Mon. Not. R. Astron. Soc.}

\def\em {\sl }

\def\eps@scaling{1.0}%
\newcommand\epsscale[1]{\gdef\eps@scaling{#1}}%
\newcommand\plotone[1]{%
 \centering 
 \leavevmode 
 \includegraphics[width={\eps@scaling\textwidth}]{#1}%
}%
\newcommand\plottwo[2]{%
 \centering 
 \leavevmode 
 \textwidth=.45\textwidth 
 \includegraphics[width={\eps@scaling\textwidth}]{#1}%
 \hfil
 \includegraphics[width={\eps@scaling\textwidth}]{#2}%
}%
\newcommand\plotfiddle[7]{%
 \centering 
 \leavevmode 
 \vbox\@to#2{\rule{\z@}{#2}}%
 \includegraphics[%
  scale=#4, 
  angle=#3, 
  origin=c 
 ]{#1}%
}%

\def\lstar{\ifmmode{\Lambda^*}\else\hbox{$\Lambda^*$}\fi} 
\def\Rop{\ifmmode{[R_{ij}]}\else\hbox{$[R_{ij}]$}\fi}

\def\Rji{\ifmmode{[R_{ji}]}\else\hbox{$[R_{ji}]$}\fi}
\def\Rstar{\ifmmode{[R_{ij}^*]}\else\hbox{$[R_{ij}^*]$}\fi}

\def\Rjistar{\ifmmode{[R_{ji}^*]}\else\hbox{$[R_{ji}^*]$}\fi}
\def\DRji{\ifmmode{[\Delta R_{ji}]}\else\hbox{$[\Delta R_{ji}]$}\fi}
\def\DRij{\ifmmode{[\Delta R_{ij}]}\else\hbox{$[\Delta R_{ij}]$}\fi}

\def\ns{\ifmmode{N_{\rm s}}          
        \else\hbox{$N_{\rm s}$}\fi}

\newcommand{\gcm}{g~cm$^{-3}$}

\begin{document}
\title{Numerical and Physical Challenges to Nebular Spectroscopy in Thermonuclear Supernovae}
\author{P. Hoeflich$^1$, E. Fereidouni$^1$, A. Fisher $^1$, T. Mera$^1$,
 C. Ashall$^2$, P. Brown$^3$, E. Baron$^4$,  J. DerKacy$^5$, T. Diamond $^6$,M. Shabandeh$^5$ , M. Stritzinger $^7$}
\address{$^1$Department of Physics, Florida State University, Tallahassee, 32306, USA}
\address{$^2$Department of Physics, Virginia Tech, Blacksburg, VA 24061, USA}
\address{$^3$Texas A\&M University, Dept. of Physics \& Astronomy, College Station, TX 77843, USA}
\address{$^4$Planetary Science Institute, 1700 E Fort Lowell Rd., Ste 106, Tucson, AZ 85719 USA}

\address{$^5$Space Telescope Science Institute, 3700 San Martin Drive, Baltimore, MD 21218-2410, USA}
\address{$^6$private astronomer}
\address{$^7$Dept. of Physics $\&$ Astronomy, Aarhus University, Ny Munkegade 120,Aarhus, Denmark}

\ead{phoeflich@fsu.edu}

\begin{abstract}Thermodynamical Supernovae (SNe~Ia) are one of the keys to high
precision cosmology and decipher the nature of the dark energy and matter.
They provide a playground for numerical astrophysical processes for a
diverse group of explosions of White Dwarf (WD) stars. At late times during the nebular phase, mid-infrared (MIR) observations are an effective tool to probe for the multi-dimensional imprints of the explosion physics of WDs and their progenitor systems. 
What we observe as SNe Ia are low-energy photons, namely light curves, and spectra detected  some days to years after the explosion. The light is emitted from a rapidly expanding envelope consisting of a low-density and low-temperature plasma with atomic population numbers far from thermodynamical equilibrium. SNe Ia are powered 
radioactive decays which produce hard X- and $\gamma$ rays and MeV leptons which are converted within the ejecta to low-energy photons. We find that the optical and IR nebular spectra depend sensitively on the proper treatment of the physical conversion of high to low energies. The low-energy photons produced by forbidden line transitions originate from a mostly optically thin envelope. However, the UV is optically thick because of a quasi-continuum formed by allowed lines and bound-free transitions even several years after the explosion. We find that stimulated recombination limits the over-ionization of high ions  with populations governed by the far UV.
The requirements to simulate nebular spectra are well beyond both 'classical' stellar atmospheres and nebulae. 
Using our full non-LTE HYDrodynamical RAdiation code (HYDRA) as a testbed, the sensitivity on the physics on 
synthetic spectra are demonstrated using observations as a benchmark. At some examples, we establish the
power of high-precision nebular spectroscopy as quantitative tool. Centrally ignited, off-center delayed-detonation near $M_{Ch}$ models can reproduce line-ratios and line profiles of Branch-normal and underluminous SNe~Ia observed with JWST. 
\end{abstract}
\begin{figure}[ht]
\includegraphics[width=0.39\textwidth]{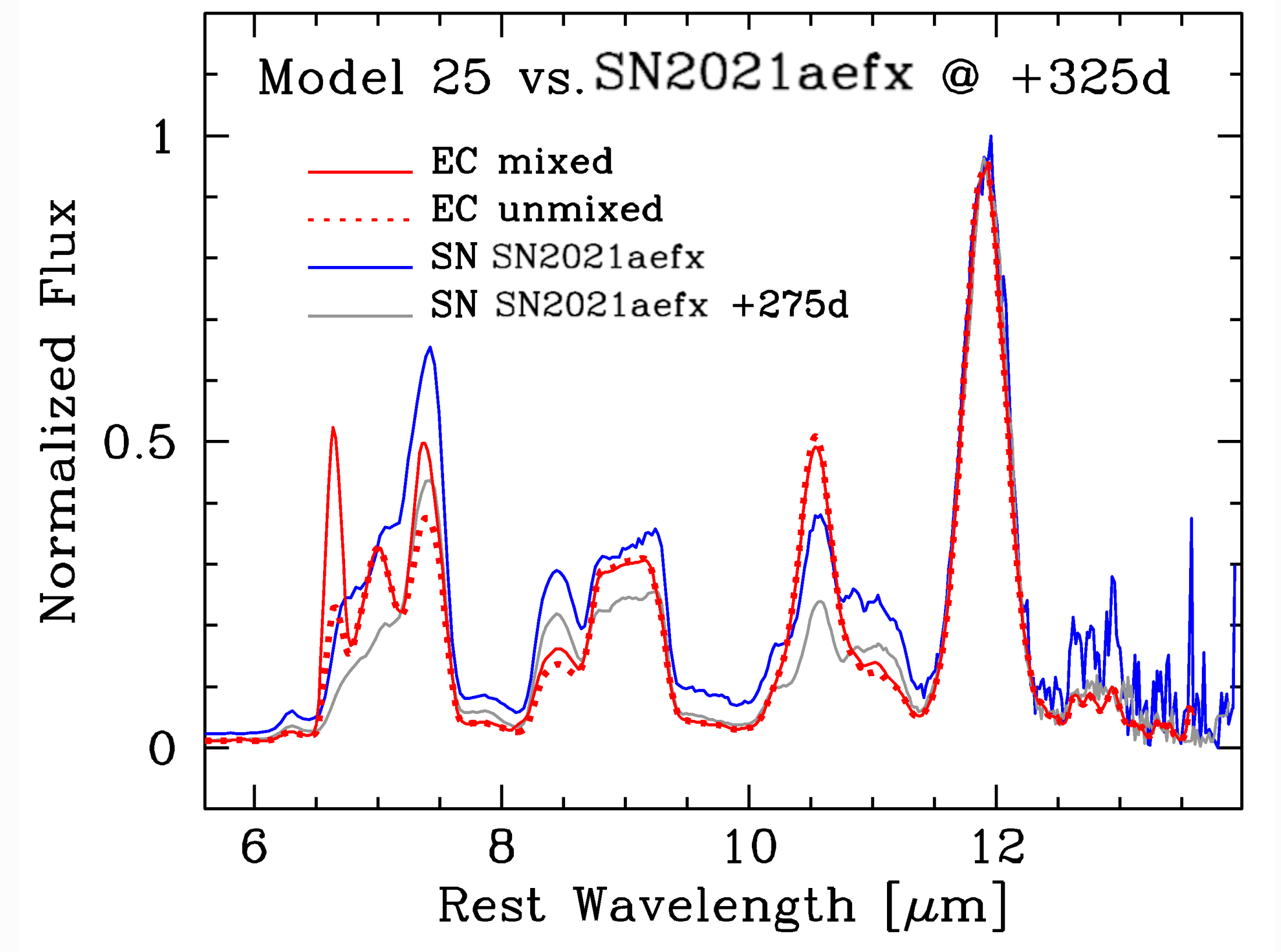}
\includegraphics[width=0.59\textwidth]{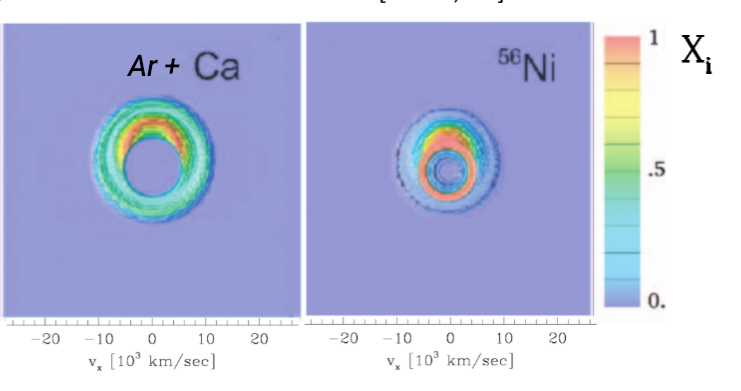}
\vskip -0.3cm
\caption{Comparison between low-resolution, mid-infrared (MIR) spectra  of the normal-bright supernovae SN2021aefx at day +275 and +325  \cite{2023ApJ...944L...3K,2023ApJ...945L...2D} obtained with MIRI/LRS at JWST  
and off-center a delayed detonation model with $\rho_c =~1\times 10^{9}$~\gcm (solid red) seen from an inclination angle of $+30^o$ \citep{2023ApJ...944L...3K}. On the right, the typical abundance of Ar and Ca distributions are given \citep{2021ApJ...922..186H}.  All spectra have been normalized using
a distance module of $32.52$~mag to bring the observed and synthetic fluxes to the same scale. The tilted feature at 9 $\mu m$ is directly related to the asymmetric [Ar] distribution with a central hole, the broad [CoIII] at 11.8 $\mu m$ reflects the $^{56}Co$ distribution, and the narrow 6.5 $\mu m$ feature can be attributed to stable $^{58}Ni$ formed in a high-density central region. Microscopic mixing can be excluded.}
\label{fig5}
\vskip -12pt
\end{figure}
\vskip -12pt
\section{Introduction}
SNe~Ia originate from the explosion of WD stars in binary or triple systems
and are key for modern cosmology and astrophysics. Advances in computational methods and time and wavelength domain observations provide new opportunities for understanding these systems.

\begin{figure*}[ht]
 \includegraphics[width=0.95\textwidth]{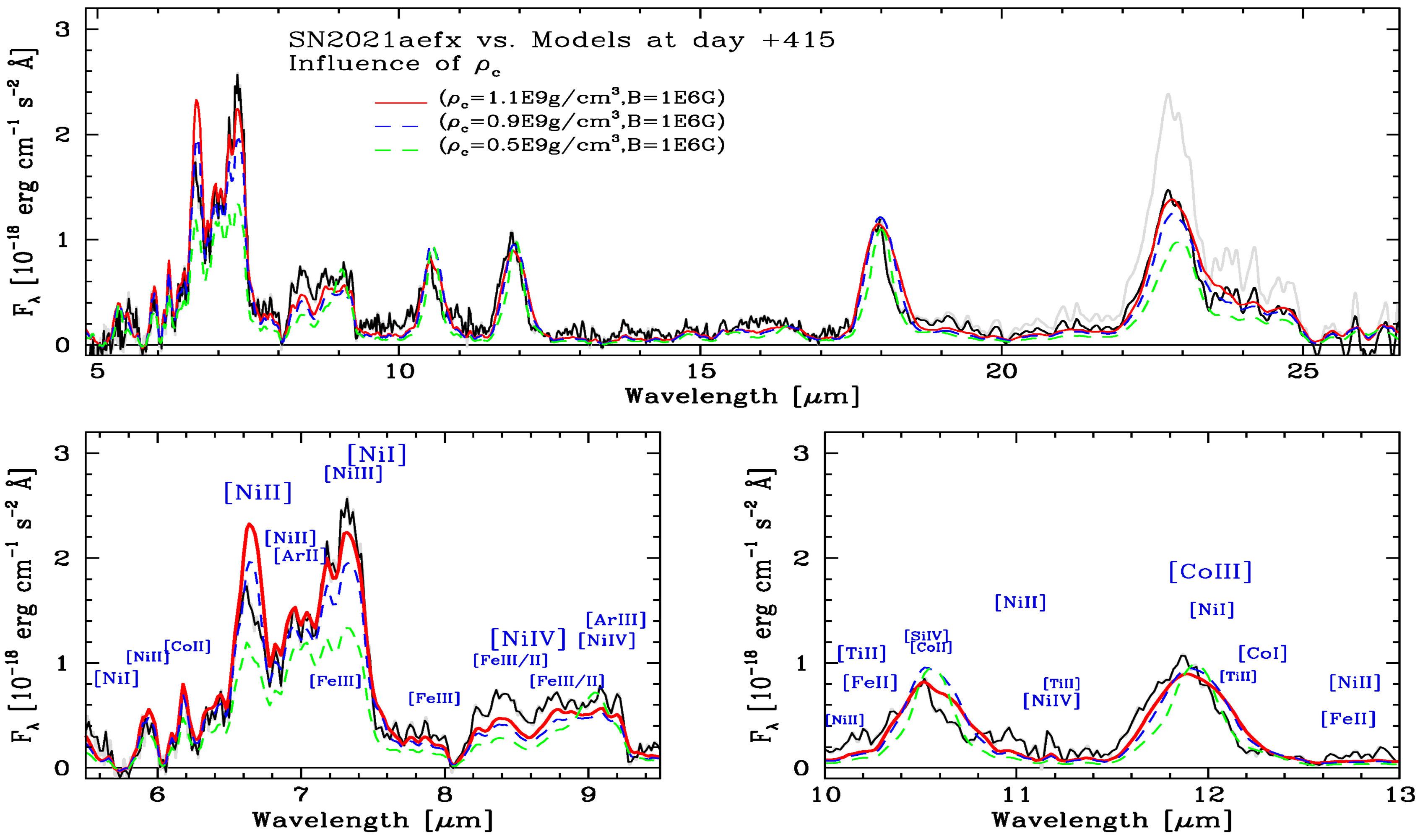}
 \vskip -5pt
  \caption{High-precision spectroscopy allows us to constrain progenitors and underlying physics.
  Comparison of the overall synthetic MIR spectra of a normal bright, off-center delayed detonation (\citep{2023MNRAS.520..560H} seen from $-30^{\circ}$),  and the \textit{JWST} MIRI/MRS  smoothed (black) and raw (gray) spectra of SN~2021aefx at $+435$ days relative after explosion. The lower plots show the zoomed regions of 5.5-9.5 $\mu m$ (left) and 10-13 $\mu m$ (right), respectively.  
    Note the sensitivity of $\rho_c$ on the overall 7 and 9 $\mu m$ features, which, mostly, is a result of the decreasing $^{58}$Ni abundance and the change in the profiles at, e.g., 11.8 vs. 10.5 $\mu m$. The best fit densities suggest a near WD mass of $\approx 1.33..1.35 M_\odot$, the narrow $^{58}Ni$ lines suggest a central ignition  \citep{hs02,khokh01,gamezo03} and exclude multiple-spot off-center ignitions suggested \citep{2013MNRAS.429.1156S,2024A&A...686A.227P}. The lack of positron transport effects is consistent with high magnetic fields \citep{Diamond2015,2021ApJ...922..186H} 
    } 
    \vskip -12pt
  \label{fig6}
\end{figure*}

From theory, SNe Ia are thermonuclear explosions of White Dwarfs (WD) and are homogeneous because nuclear physics determines the WD structure and explosion, so-called stellar amnesia \citep{2006NuPhA.777...579H}. The total amount of burning gives the total energy production. The light curves are determined by the amount of radioactive $^{56}Ni$. . Thus, a wide range of explosion scenarios produce very similar spectra and light curves, mostly masking complex 3D physical processes and the initial boundary conditions imposed by the progenitor system. To reveal the underlying physics requires high-precision spectroscopy observations and radiation hydro or magneto-hydrodynamical simulations of matching accuracy. Though both full spherical and multi-dimensional simulations allow high-precision synthetic light curves and spectra in agreement with observations,  mixing dimensionality for hydro and transport simulations may result in severe spectral artifacts. Angle averaging of asymmetric abundances and using spherical transport induces microscopic mixing through the atomic rate equations, significantly altering the synthetic spectra and their interpretation \cite{2015ApJ...806..107D}, and see Fig. \ref{fig5}. The current status of spectral analysis is shown in
Fig.\ref{fig6}.
 
However, despite the advances in modeling efforts, it became further apparent that we still miss essential physical effects, such as the influence of the magnetic fields on thermonuclear runaway, flame propagation, and cellular instabilities.
From the simulations, the key questions remaining are: What accuracy in simulations is required to probe underlying physics by modern observations? Here, we want to focus on the coupling of the atomic statistical equations and the radiation transport with a focus on nebular spectra with our code HYDRA as a testbed
and observations as a benchmark.
\section{HYDRA as a testbed}

Our HYDrodynamical RAdiation (\textit{HYDRA}) code consists of physics-based
modules to provide a solution for the  nuclear networks, 
the statistical equations needed to determine the atomic level population, 
the equations of state,  the opacities,  and the hydrodynamics (without "Adaptive Mesh Refinement``), front tracking schemes for the nuclear flame,  
and radiation transport problems via Variable Eddington solvers and Monte-Carlo closure relation, including the energy deposition by high-energy photons and particles (see \citep{2024JPhCS2742a2024H,H90,2003ASPC..288..185H,2017ApJ...846...58H,2021ApJ...922..186H}, and references therein).  

\begin{figure}[ht]
\includegraphics[width=0.51\textwidth]{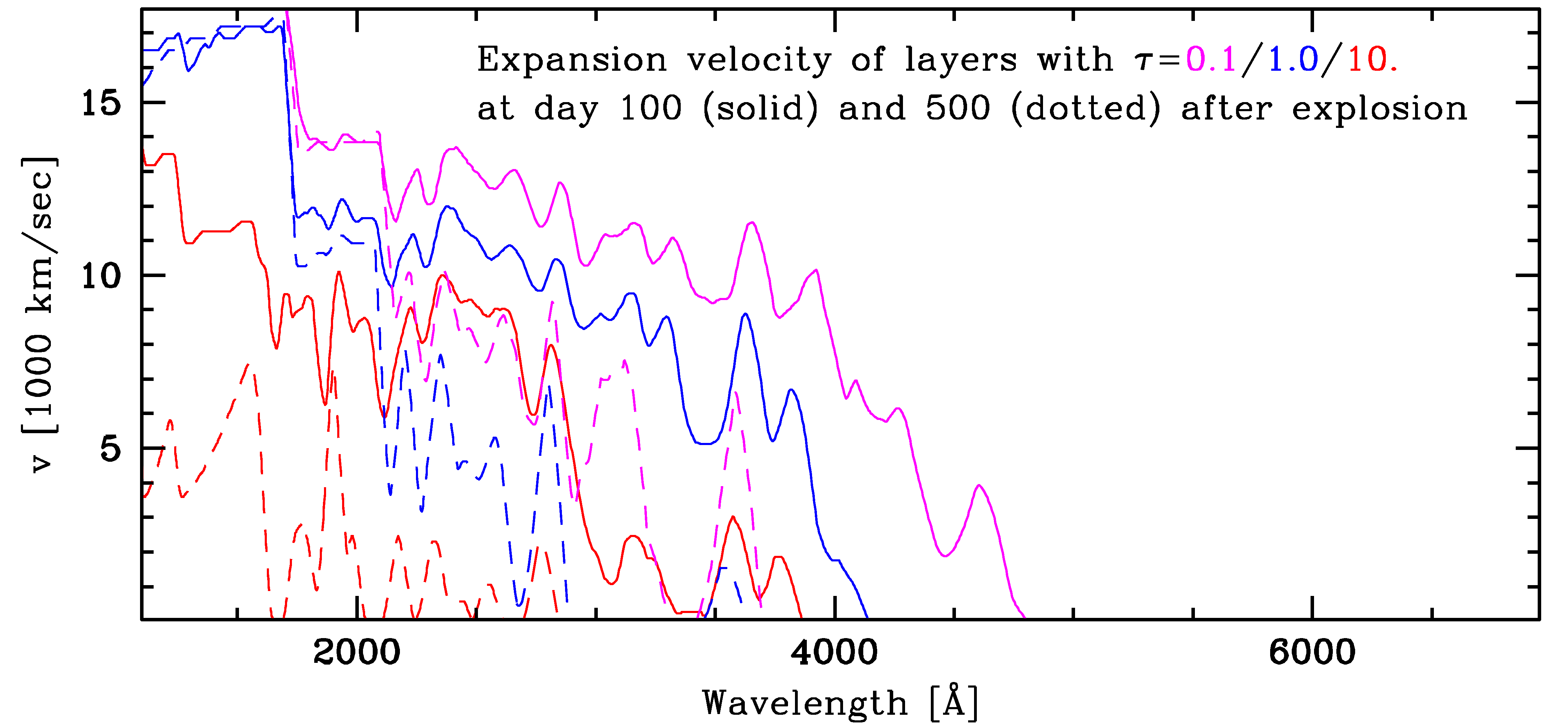} 
\includegraphics[width=0.45\textwidth]{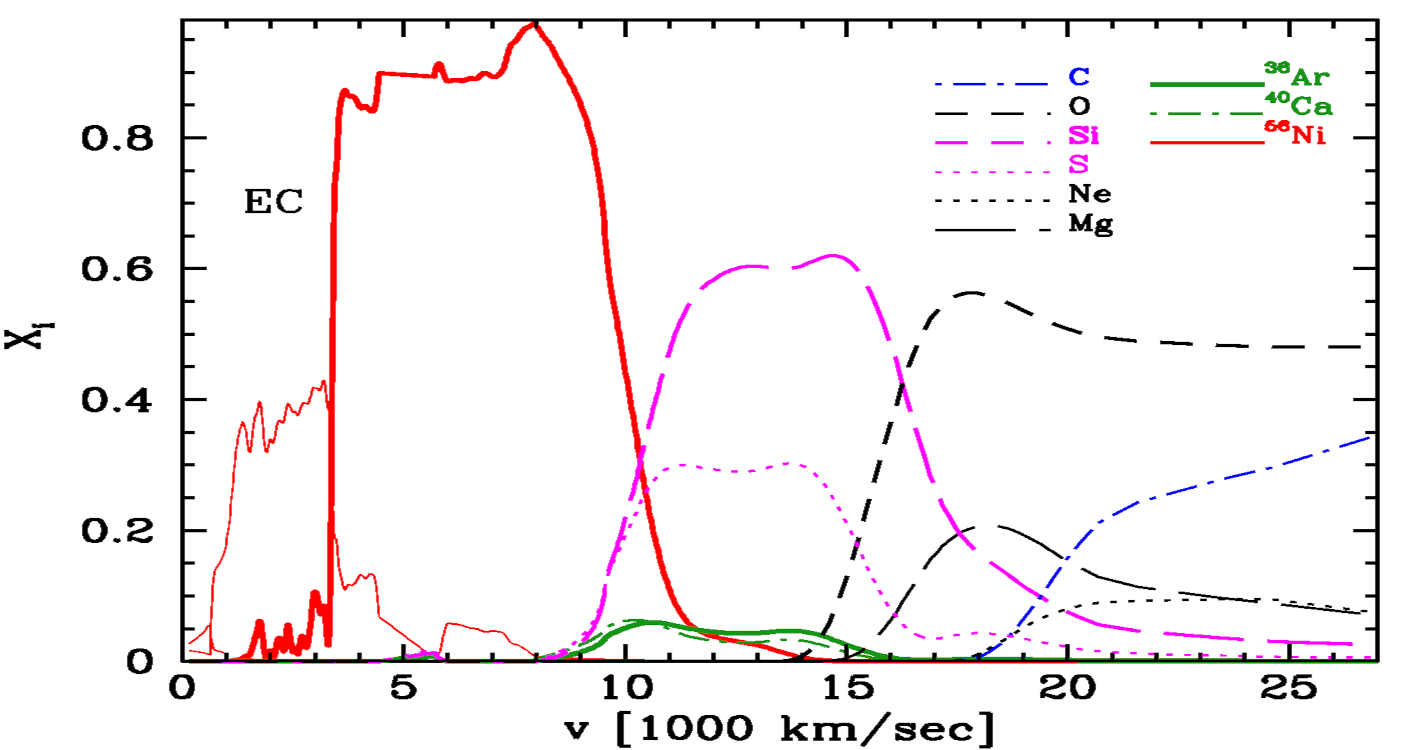}
\vskip -0.3cm
\caption{Velocity as a function of wavelength at which the optical depth reaches 0.1, 1., and 10 for a normal-bright off-center delayed detonation model during the nebular phase at about 100 and 500 days after the explosion. To link the velocity to the abundance layers, the angle averaged abundance structure as a function of expansion velocity is given and used to analyze JWST data of SN2021aefx \citep{2023ApJ...944L...3K,2024ApJ...975..203A}. Spectra are dominated by iron-group elements, and those transitions are
'pumped' by redistribution from the UV and X-ray wavelength range, setting apart qualitative SN spectroscopy from the classical theory of nebular.
}
\label{fig1}
\end{figure}


 The individual numerical and physical modules are coupled explicitly. In this contribution, we will focus on 
 the nebular phase, the ejecta is optically thin at optical wavelengths and beyond. However, even several years after the explosion, the envelope is optically thick in the UV and shorter wavelengths, and MeV photons and leptons provide the energy input. Both aspects require codes and methods different from those commonly used for stellar atmospheres and 'classical' optically thin nebulae. Specifically, we will present our method used a) for coupling radiation transport and atomic statistical equations and b) for coupling non-thermal sources with the low-energy atomic rate equations. 
 As a testbed, we use explosions of near Chandrasekhar mass WDs ($M_{Ch}$) \citep{khok89,h95,hk96} for normal-bright and subluminous SNe~Ia where the flame starts close to the center and transitions to a detonation. 
 HYDRASSON, HYDRA'S SON, was created as part of the PhD thesis of A. Fisher as a computational fast alternative (factor 100...500) for late-time spectra by stripping many advanced features to study the sensitivity on, e.g. of the atomic data and physical approximations, for given explosion models. Although using many modules of HYDRA, the acceleration was achieved by assuming spherical geometry, a simplified treatment of atomic physics, and assuming that elements are pumped by radioactive energy according to their baryon number. Currently, the second author identifies the physics needed, without sacrificing speed, by implementing proper bound-free opacities and using the Spencer-Fano equation to be calibrated by the Monte Carlo (MC) of HYDRA in the future (see Fig. \ref{fig3}). A fast code is needed to process a large number of supernovae and in light of a large number of variables. The goal is to find fast spherical solutions and to use 3D HYDRA for the final analysis of well-observed objects.


\subsection{Atomic models and data}
We construct detailed atomic models depending on the application and phase of the supernovae and reduce the elements, number of ionization stages, and excitation levels accordingly. We omit solving the rate equations for elements in zones with particle abundances less than $^10{-5}$. 
Detailed atomic models are used for the ionization stages I-V for C, O, Ne, Mg, Si, S, Cl, Ar, Ca, Sc, Ti, V, Cr, Mn, Fe, Co, Ni. 
The atomic models and line lists are constructed from the database for both allowed and forbidden bound-bound transitions \citep{vanHoof2018} (see {\url{https://www.pa.uky.edu/~peter/newpage/}}), supplemented by additional forbidden lines from previous analysis \cite{Diamond2015,2021ApJ...923..210H} and, for weak forbidden lines without mid-IR transition probabilities, by a comparison between synthetic and observed spectra \citep{2024ApJ...975..203A}. The photon and collisional cross-sections for bound-free transitions, the charge transfer and Auger effects are taken from \citep{1996ApJ...465..487V} (see\url{http://www.pa.uky.edu/~verner/atom.html},\url{https://www.pa.uky.edu/~verner/photo.html}). The importance of the atomic bound-free rates is shown in Fig. \ref{fig3}. Using detailed bound-free rates (green line, lower right) allows us to reproduce the [CoIII](11.8 $\mu m$)/[CoII](10.3 $\mu m$) line ratio. In contrast, the ratios between features are off by more than a factor of two (lower left plot, magenta line).

The detailed atomic models are constructed as follows: a) The level scheme is based on all levels; b) Superlevels are created by merging for the rate equations if $\Delta E_{i,j}$ are smaller than the kinetic energy. For ions with low abundances or not apparent in the spectra analyzed, levels $\Delta E \approx 0.1eV$ are merged for computational efficiency. 
Note that the atomic levels $E_i$ are based on levels with and without known transition probabilities because collisions are crucial to avoid the IR catastrophe \citep{axelphd80,fransson94}, and because of its apparent absence in observations. This is important to simulate the effective critical density by allowing electrons in levels to cascade down in small steps. Collisional de-excitation rates scale with $\exp(-\Delta {E({ij}\over kT)}$). Note that allowed transitions must be included during the nebular phase because they form an optically thick quasi-continuum
in the UV up to several years after the explosion (Fig. \ref{fig1}), and UV dominates the ionization balance by trapping UV photons and operating incomplete Rosseland Cycles to transfer high-energy photons to the optical and IR \citep{2021ApJ...923..210H}, and see below.

\subsection{Iterative schemes employed to couple rate and radiation transport}

We solve the full set of statistical equations to determine the population
density of atomic states, $n_i$, where $i$ is shorthand for the excitation state $i$ out of $js$ excitation states for the ionization state $k$ of element $el$. In the following, the index {\sl el} is omitted, but the charge-equation couples the level populations of elements. Time-dependence is solved for bound-free equations \citep{h95}
. The level populations are given by the $el \times k
 \times j$ equations,

$$
{\partial n_i \over \partial t} + \nabla (n_{i} {\bf v}) =
\sum_{i\ne j}
 (n_{j} P_{ji} - n_{i} P_{ij}) + n_{k'} (R_{k'i} + C_{k'i}) - n_{i} (R_{ik'} + 
C_{ik'}).~~~~~~~~~~~~~~~(1)
$$

Here the rate, $P_{ij}$, is the sum of the bound-bound radiative, $R_{ij}$, and collisional,
$C_{ij}$, processes between levels i and j within a specific ion;  $k'$ stands for
all transitions from all bound levels to higher ionization states.
Note that the total opacity $\kappa$, emissivity $\eta$, and the source function $S$ at a given frequency $\nu$ are the sum over all elements, ionization stages, and levels (see eq. B13 in \cite{2021ApJ...923..210H}).

The radiative rates between a lower and upper level i and j,
respectively, are given by
$$ 
R_{ij} = 4 \pi \int
{\alpha_{ij} (\nu) \over h \nu} J_\nu d\nu 
 {\rm  + P_{\gamma,lepton} ~~and~~ } R_{ji} = 4 \pi  \int
{\alpha_{ij}(\nu) \over h \nu }
[{{2 h \nu^3 \over c^2} + J_\nu} e^{-{h \nu \over k T}}] d\nu  
~~~~~~~~~(2)
$$ 
with the cross-section $\alpha_{ij}$ $P_{\gamma,lepton}$  being the rates
induced by high-energy particles and photons (see below).

For transition between a lower and an upper level l and u,
respectively. The departure coefficients $b$ are defined by the population number of each level relative to the next higher ionization stage. 
 The explicit form of the source function $S$ and formal solution of the radiation field $J_\nu$  are given by
$$ 
\bar S_{lu} (\nu) = {2 h \nu^3 \over c^2} {1 \over ({b_l \over b_u} e^{h \nu 
/ k T} - 1)}, {\rm ~~~and~~~ } \vec{J_\nu}= \Lambda{_\nu}  \vec{S_\nu} ~~~~~~~~~~~~~~~~~~~~~~~~~~~~~~~~~~~~~~~~(3)
$$ 
with $\Lambda$ being the 'unknown' radiation-transport matrix in explicit form, and $\vec{J_\nu}$ the mean intensity at every grip point, and the departure coefficients $b_i$ being defined as the ratio between the non-LTE level population relative to the LTE population normalized to the next higher ion. Note that the stimulated recombination rates can be significantly boosted for $b_l < 1$. The rise of the recombination
puts a limit on the over-ionization by non-thermal processes (see below).

\subsection{Coupling the statistical and radiative transport equation for low-energy photons}

 In the partially or fully optical thin case, the classical nebular regime, a consistent solution of equations 3 can obtained using $S^{(m-1)}$  to obtain $J^{m}$ with $m$ being the model iteration. This approach
 is used at wavelengths for which the envelope is semi-transparent.
 
In the optically thick case, the local $J$ produces $S$, and the scheme does not converge.
 This problem is well-known with the so-called 'accelerated` $\Lambda$ iteration where $\Lambda$ is substituted by approximation for $\Lambda^*$ \cite{scharmer84,1986JQSRT..35..431O,hillier90,2003ASPC..288..185H,2003aspc..288...51h}, which drives the convergence.

Case A: If the  transition in consideration dominates the total
source function, we use the approach
 $S_{\nu} \approx S_{ij} (\nu) $ and $J^m \approx \Lambda S^{m-1}+ \Lambda^* (S^{(m)}-S^{(m-1)})$, to
obtain "first" order corrections to the
rates and substitute $J$ in the rate equations. Note that the correction terms below are scaled according to the specific relative to the total opacity and use the concept of leading elements as discussed at the end.

 If we substitute the radiative rates
in the statistical equations, we get
$$ 
\eqalign{
&b_i (C_{i j} +  R_{i j} + \sum_{j \ne i} P_{i j}) - b_j (C_{ij} +
 R_{ji} + \sum_{j \ne i} P_{j i}) = \cr
&b_j \int {\alpha_{ij}(\nu) \over h \nu}\Lambda_{\nu}^*
(S_{ij}^{(m)}(\nu)-
S_{ij}^{(m - 1)}(\nu))
e^{-{h \nu \over k T}} d \nu - b_i \int {\alpha_{ij} \over h \nu}
\Lambda_{\nu}^* (S_\nu^{(m)} - S_{ij}^{(m - 1)}(\nu)) d \nu 
 \cr}.
~~~~~~(4)
$$ 

The explicit form of the source functions, eq. 3,
allows us to substitute the departure coefficient of the
lower level. After some simple transformations
 one ends up with the expression
for the first-order correction of the rates
$$ 
R_{ij}^{corr} = 0, {\rm ~~and~~ }
R_{ji}^{corr} = 4 \pi \int {\alpha_{i j} (\nu) \over h
\nu} \Lambda_{ij}^*(\nu) {(S_{ij}^{(m-1)}(\nu)
 - S^{(m)}_{ij} (\nu))\over S_{
ij }^{(m)}(\nu)} d \nu
~~~~~~~~~~~~~~~~~(5)
$$ 

Case B: If 
$S_\nu/\chi_\nu \gg S_{ij}/\chi_{ij}$, the correction terms to the rates $R$ obtained from
the previous iterations can be written as
$$ 
R_{ij}^{corr} =
 4 \pi \int {\alpha_{i j} (\nu)\over h \nu} \Lambda_\nu^*
  (S^{(m)}_\nu - S^{(m
-
1)}_\nu) d\nu  {\rm~~ and~~}
R_{ji}^{corr} =
 4 \pi \int {\alpha_{i j}(\nu) \over h \nu}
 \Lambda_\nu^*
 (S^{(m)}_\nu - S^{(m
-
1)}_\nu)
e^{{-h \nu \over k T}} d~ \nu. ~~(6)
$$ 

To accelerate the convergence, a second-order extrapolation in $m$
is employed \citep{H90}.

\subsection{Coupling between different elements and ions}
 In principle, coupling between elements occurs via charge conservation.
 For the recombination, both stimulated and spontaneous emission to the ground and excited states are essential. 
 However, for low temperatures, the dynamic range of values for $J$ relevant for high ions exceeds the range of numbers. Radiation transport would give zero intensities. For cold plasmas powered by radioactive decay,
 high ionization stages are important, and most of the transitions are in optically thick far-UV and X-rays. In practice, for sizeable optical depth ($\tau \ge 10$ within a grid), an incomplete, stationary Rosseland cycle is employed to calculate the flow to long wavelengths, assuming local radiation fields effectively 'feed' the levels with optically thin transitions as a loss term for the cycle with non-thermal, high energy rates as gain terms. The result is an under-population of the lower ionization ($b_i<1$,  see eq. 2), boosting the stimulated emission rates and force recombination. In literature, this effect is sometimes mimicked by increasing the recombination rate assuming high-density clumps \citep{Wilk2020}, or artifically increasing the recombination rates by a free factor \citep{Shingles2020}. 

 Using the rate equation, one gets the following expression for the redistribution between levels (and a similar expression for bound-free transitions) \citep{H90}. It allow us to accelerate the redistribution
 between elements, ions, and from the X-ray and UV to optical wavelengths and beyond.
$$  
\bar S_{lu} = {\bigl(\int \phi_\nu J_\nu  + (\varepsilon^\prime +
\theta)
B_\nu\bigr) \over \bigl(1 + \varepsilon^\prime + \varsigma \bigr)
}
~{\rm with}~
 \int \phi_\nu d \nu = 1 \eqno
~~~~~~~~~~~~~~~~~~~~~~(7)
$$ 
with the inverse lifetime $A_{ij}$, one obtains
$$ 
\varsigma = {{a_2 a_3 - {g_l \over g_u} a_1 a_4} \over A_{ul} (a_2
+
a_4)}, 
\theta = {n^*_1 a_l a_4 (1 - e^{-h \nu /k T}) \over (A_{ul} n^*_u (a_2
+ a_4))}, 
\\
\varepsilon^\prime = C_{ul} (1 - e^{- h \nu / k T}) / A_{ul} 
~~~~~~~~~~~~~~~~~~~~~~~~~~(8)
$$ 
with
$$
a_1 = R_{lk} + C_{lk}+\sum_{i < l} {n_i^* \over n_l^*} R_{li} - {n_i \over n_l}
R_{il} + \sum_{l < j \ne u} {(1 - {b_j \over b_l} C_{lj})},
$$
$$
a_2 = n_l^* (R_{kl} + C_{lk}) + \sum_{l < j \ne u} b_j n_l^* R_{jl} -
n_l R_{lj} + \sum_{i < l} n_i C_{il} ({1 - {b_l \over b_i}})
$$
$$
a_3 = R_{uk} + C_{uk}+\sum_{u >i \ne  l} {n_i^* \over n_u^*} R_{ui} - {n_i \over n_u}
R_{iu} + \sum_{u < j } ({1 - {b_j \over b_u}}) C_{uj}
$$
$$
a_4 = n^*_u (R_{ku} + C_{uk}) + \sum_{u < j} b_j n^*_u R_{ju} - n_u
R_{uj} + \sum_{u > i \ne l} n_i C_{iu} ({1 - {b_u \over b_i}}).
$$

For nebular spectra and the optically thick case and when a transition dominates, we use the fraction of the dominant source function of any element or ion for $S_\nu$ as the driving term for all coefficients of eq. 8. We solve the rate equation in order of abundance at a particular computational cell.

\begin{figure}[ht]
\includegraphics[width=0.95\textwidth]{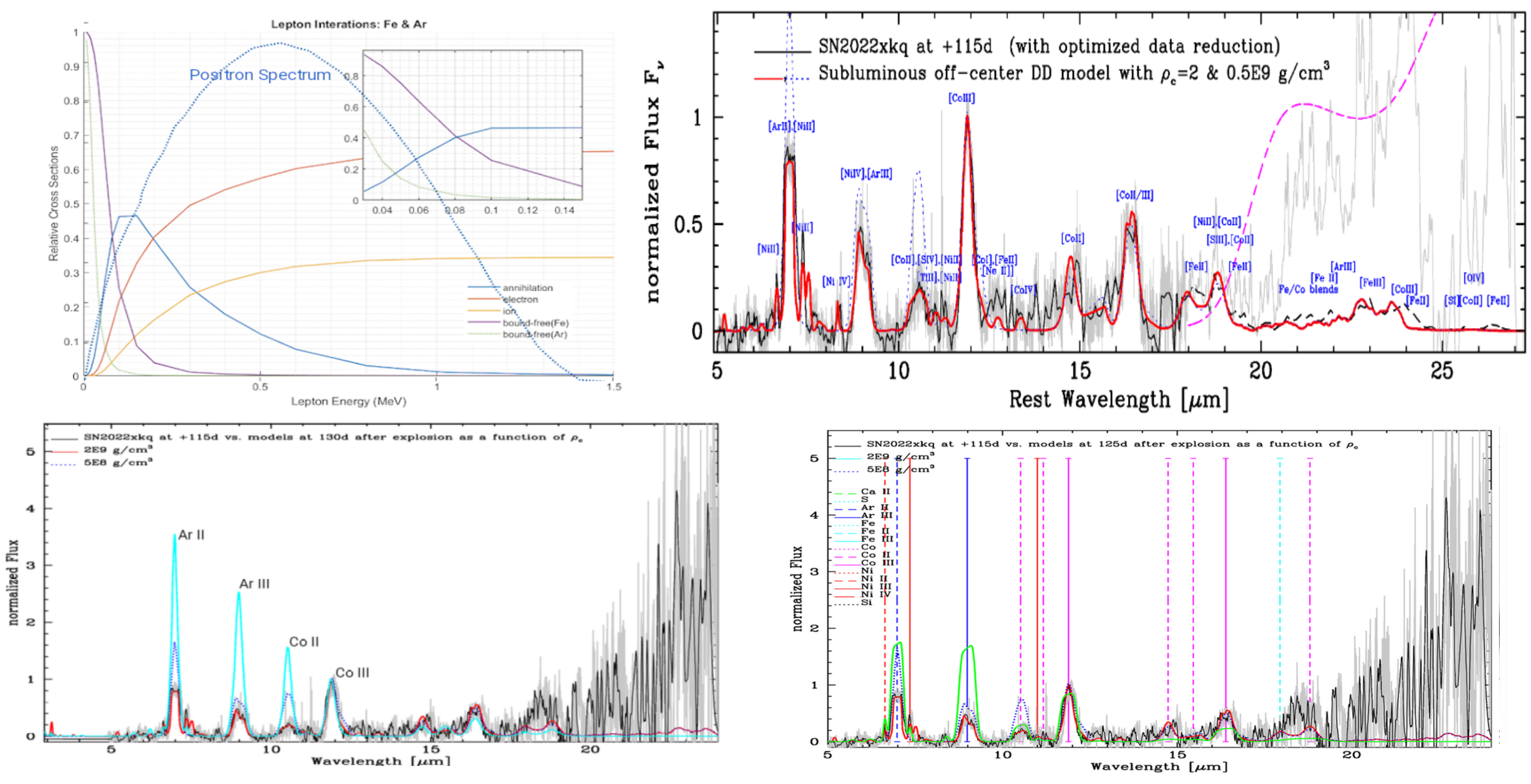}
\vskip -0.3cm
\caption{Influence of the treatment of non-thermal particles and low-energy bound-free cross sections on nebular spectra. {\sl Upper left:} Energy-distribution of positron originating from the $\beta^+$-decay of $^{56}Co \rightarrow ^{56}Fe$, and normalized cross sections of positron-electron annihilation, lepton-lepton, lepton-ion interaction, and as an example, ionization from the inner s-electrons of Fe and Ar, respectively. For example, the 
comparison is given between the MIR spectra some 115days after maximum of the subluminous SN2022xkq  and \cite{2023ApJ...944L...3K} obtained with MIRI/LRS at JWST
and models with $\rho_c =~2\times 10^{9}$~\gcm (solid red) and $5 \times 10^{8}$~\gcm (dotted blue) using detailed bound-free cross-sections and a full Monte Carlo approach to the energy transfer (case a, see text). Overall, the synthetic and observed spectra show good agreement.
 {\sl lower left:}  Same but including the synthetic spectrum (magenta) under the assumption of a hydrogen-like approximation and high-energy input proportional to the electrons per element (case b). 
 {\sl lower right:} Same but including the detailed bound-free cross sections and an energy deposition close
 to the solution for the Spencer-Fano equation (case b). The latter provides a significant improvement compared to the simulation on the lower left, but both fail to produce correct Ar to Co line ratios (see text).
}
\label{fig3}
\end{figure}

\section{Ionization, $\gamma $-rays and non-thermal leptons}

The energy deposition by non-thermal particles, $\gamma $-rays and non-thermal leptons, and enters the rate equations (eq.2)  via non-thermal ionization \cite{KF92} and inelastic ion scatter balanced by the recombination processes. For the nebular phase, we assume stationarity. The nuclear energy input per time is balanced by the local flux \citep{hkm92,h04,penney14,2021ApJ...923..210H}.

Detailed simulations of nebula spectra and ionization balances were started by \cite{Axelrod1980}. SNe light curves are powered by radioactive decay $^{56}Ni\rightarrow^{56}Co\rightarrow ^{56}Fe$. This complex process is treated in various approximations, namely a) by relating the number of ionization processes directly to the electron per element   (e.g. \citep{hillier90,Kasen2006,Sim10,2023ApJ...944L...3K}), b) by treating the non-thermal excitation and ionization using the Spencer-Fano equations \citep{Spencer54} (e.g. \citep{2015ApJ...814L...2F,2017ApJ...845..176B,Shingles2020,Wilk2018,Wilk2020,Shingles2020}, or c) by using detailed Monte Carlo simulations as in HYDRA which allows to follow the process in detail \cite{2021ApJ...922..186H,2023ApJ...944L...3K,2024ApJ...975..203A}.

Most of the results published show significant shortcomings, in particular in the MIR and when using stellar codes. A  discussion of the physical processes involved and evaluation of the differences of codes is beyond  beyond the scope of this paper. Thus, we used the HYDRA framework because it can reproduce MIR spectra (right upper plot of Fig. \ref{fig3}).
The corresponding levels are filled by a series of radiative transitions corresponding to the energy difference between levels between non-valence electrons and Auger transitions, using the probability of multiple electron emission \cite{Kaastra1993}.
The impact of assumption a) (and Hydrogen approximation for bound-free) is seen in the lower left. The ratio between the [Co II/CoIII] is off, and too much energy is pumped into the Ar features. Using HYDRASSON with updated bound-free rates and energy input close to the explicit solution of the Spencer-Fano equation provides a significant improvement; namely, the CoII/CoIII ratio is corrected, but still, Ar is pumped in excess. The reason becomes evident from the  cross-sections. Non-thermal electrons are cascading down; however, the ionization effect is from inner-shell electrons. Namely, the cascade first feeds into the iron-group elements (e.g. Fe/Co/Ni), and only the 'left-overs' can power intermediate mass elements such as Ar and Ca.

\noindent
{\sl Acknowledgments:} Ph. acknowledges support by the NSF grants AST-1715133 and AST-2306395, and NASA
grants JWST-GO-02114, GO-02122, GO-03727, GO-004217, GO-04436, GO-4522, GO-5057, GO-05290, GO-06023, GO-6677,  HST-16190.
Thanks to all members of our JWST and SPECPOL/VLT teams.

\providecommand{\newblock}{}


\end{document}